\documentclass[aps,prd,twocolumn,amsmath,amssymb,amsfonts,nofootinbib,superscriptaddress,altaffilletter]{revtex4-1}
\usepackage{graphicx}
\usepackage{hyperref}
\hypersetup{
  bookmarksopen=true
}
\usepackage{amssymb}
\usepackage{amsmath}
\usepackage{color}
\usepackage{supertabular}
\usepackage{dcolumn}
\usepackage{wasysym}
\usepackage{acronym}

\begin{document}

\title{Search for continuous gravitational waves from neutron stars in globular cluster NGC~6544}

\begin{abstract}
We describe a directed search for continuous gravitational waves in data from the sixth initial LIGO science run. The target was the nearby globular cluster NGC~6544 at a distance of $\approx 2.7$ kpc. The search covered a broad band of frequencies along with first and second frequency derivatives for a fixed sky position. The search coherently integrated data from the two LIGO interferometers over a time span of 9.2 days using the matched-filtering $\mathcal{F}$-statistic. We found no gravitational-wave signals and set 95\% confidence upper limits as stringent as $6.0 \times 10^{-25}$ on intrinsic strain and $8.5 \times 10^{-6}$ on fiducial ellipticity. These values beat the indirect limits from energy conservation for stars with characteristic spindown ages older than 300 years and are within the range of theoretical predictions for possible neutron-star ellipticities. An important feature of this search was use of a barycentric resampling algorithm which substantially reduced computational cost; this method will be used extensively in searches of Advanced LIGO and Virgo detector data.
\end{abstract}




\author{%
B.~P.~Abbott,$^{1}$  
R.~Abbott,$^{1}$  
T.~D.~Abbott,$^{2}$  
M.~R.~Abernathy,$^{3}$  
F.~Acernese,$^{4,5}$ 
K.~Ackley,$^{6}$  
C.~Adams,$^{7}$  
T.~Adams,$^{8}$ 
P.~Addesso,$^{9}$  
R.~X.~Adhikari,$^{1}$
% ...
\\
\medskip
(LIGO Scientific Collaboration and Virgo Collaboration) 
\\
\medskip
and S. Sigurdsson$^{73}$
\\
\medskip
{{}$^{*}$Deceased, March 2016. {}$^{\dag}$Deceased, May 2015. {}$^{\ddag}$Deceased, March 2015. }%
}\noaffiliation
\affiliation {LIGO, California Institute of Technology, Pasadena, CA 91125, USA }
\affiliation {Louisiana State University, Baton Rouge, LA 70803, USA }
\affiliation {American University, Washington, D.C. 20016, USA }
\affiliation {Universit\`a di Salerno, Fisciano, I-84084 Salerno, Italy }
\affiliation {INFN, Sezione di Napoli, Complesso Universitario di Monte S.Angelo, I-80126 Napoli, Italy }
\affiliation {University of Florida, Gainesville, FL 32611, USA }
\affiliation {LIGO Livingston Observatory, Livingston, LA 70754, USA }
\affiliation {Laboratoire d'Annecy-le-Vieux de Physique des Particules (LAPP), Universit\'e Savoie Mont Blanc, CNRS/IN2P3, F-74941 Annecy-le-Vieux, France }
\affiliation {University of Sannio at Benevento, I-82100 Benevento, Italy and INFN, Sezione di Napoli, I-80100 Napoli, Italy }
\affiliation {Albert-Einstein-Institut, Max-Planck-Institut f\"ur Gravi\-ta\-tions\-physik, D-30167 Hannover, Germany }
\affiliation {Nikhef, Science Park, 1098 XG Amsterdam, The Netherlands }
\affiliation {LIGO, Massachusetts Institute of Technology, Cambridge, MA 02139, USA }
\affiliation {Instituto Nacional de Pesquisas Espaciais, 12227-010 S\~{a}o Jos\'{e} dos Campos, S\~{a}o Paulo, Brazil }
\affiliation {INFN, Gran Sasso Science Institute, I-67100 L'Aquila, Italy }
\affiliation {INFN, Sezione di Roma Tor Vergata, I-00133 Roma, Italy }
\affiliation {Inter-University Centre for Astronomy and Astrophysics, Pune 411007, India }
\affiliation {International Centre for Theoretical Sciences, Tata Institute of Fundamental Research, Bangalore 560012, India }
\affiliation {University of Wisconsin-Milwaukee, Milwaukee, WI 53201, USA }
\affiliation {Leibniz Universit\"at Hannover, D-30167 Hannover, Germany }
\affiliation {Universit\`a di Pisa, I-56127 Pisa, Italy }
\affiliation {INFN, Sezione di Pisa, I-56127 Pisa, Italy }
\affiliation {Australian National University, Canberra, Australian Capital Territory 0200, Australia }
\affiliation {The University of Mississippi, University, MS 38677, USA }
\affiliation {California State University Fullerton, Fullerton, CA 92831, USA }
\affiliation {LAL, Univ. Paris-Sud, CNRS/IN2P3, Universit\'e Paris-Saclay, Orsay, France }
\affiliation {Chennai Mathematical Institute, Chennai 603103, India }
\affiliation {Universit\`a di Roma Tor Vergata, I-00133 Roma, Italy }
\affiliation {University of Southampton, Southampton SO17 1BJ, United Kingdom }
\affiliation {Universit\"at Hamburg, D-22761 Hamburg, Germany }
\affiliation {INFN, Sezione di Roma, I-00185 Roma, Italy }
\affiliation {Albert-Einstein-Institut, Max-Planck-Institut f\"ur Gravitations\-physik, D-14476 Potsdam-Golm, Germany }
\affiliation {APC, AstroParticule et Cosmologie, Universit\'e Paris Diderot, CNRS/IN2P3, CEA/Irfu, Observatoire de Paris, Sorbonne Paris Cit\'e, F-75205 Paris Cedex 13, France }
\affiliation {Montana State University, Bozeman, MT 59717, USA }
\affiliation {Universit\`a di Perugia, I-06123 Perugia, Italy }
\affiliation {INFN, Sezione di Perugia, I-06123 Perugia, Italy }
\affiliation {European Gravitational Observatory (EGO), I-56021 Cascina, Pisa, Italy }
\affiliation {Syracuse University, Syracuse, NY 13244, USA }
\affiliation {SUPA, University of Glasgow, Glasgow G12 8QQ, United Kingdom }
\affiliation {LIGO Hanford Observatory, Richland, WA 99352, USA }
\affiliation {Wigner RCP, RMKI, H-1121 Budapest, Konkoly Thege Mikl\'os \'ut 29-33, Hungary }
\affiliation {Columbia University, New York, NY 10027, USA }
\affiliation {Stanford University, Stanford, CA 94305, USA }
\affiliation {Universit\`a di Padova, Dipartimento di Fisica e Astronomia, I-35131 Padova, Italy }
\affiliation {INFN, Sezione di Padova, I-35131 Padova, Italy }
\affiliation {CAMK-PAN, 00-716 Warsaw, Poland }
\affiliation {University of Birmingham, Birmingham B15 2TT, United Kingdom }
\affiliation {Universit\`a degli Studi di Genova, I-16146 Genova, Italy }
\affiliation {INFN, Sezione di Genova, I-16146 Genova, Italy }
\affiliation {RRCAT, Indore MP 452013, India }
\affiliation {Faculty of Physics, Lomonosov Moscow State University, Moscow 119991, Russia }
\affiliation {SUPA, University of the West of Scotland, Paisley PA1 2BE, United Kingdom }
\affiliation {University of Western Australia, Crawley, Western Australia 6009, Australia }
\affiliation {Department of Astrophysics/IMAPP, Radboud University Nijmegen, P.O. Box 9010, 6500 GL Nijmegen, The Netherlands }
\affiliation {Artemis, Universit\'e C\^ote d'Azur, CNRS, Observatoire C\^ote d'Azur, CS 34229, Nice cedex 4, France }
\affiliation {Institut de Physique de Rennes, CNRS, Universit\'e de Rennes 1, F-35042 Rennes, France }
\affiliation {Washington State University, Pullman, WA 99164, USA }
\affiliation {Universit\`a degli Studi di Urbino ``Carlo Bo,'' I-61029 Urbino, Italy }
\affiliation {INFN, Sezione di Firenze, I-50019 Sesto Fiorentino, Firenze, Italy }
\affiliation {University of Oregon, Eugene, OR 97403, USA }
\affiliation {Laboratoire Kastler Brossel, UPMC-Sorbonne Universit\'es, CNRS, ENS-PSL Research University, Coll\`ege de France, F-75005 Paris, France }
\affiliation {Carleton College, Northfield, MN 55057, USA }
\affiliation {Astronomical Observatory Warsaw University, 00-478 Warsaw, Poland }
\affiliation {VU University Amsterdam, 1081 HV Amsterdam, The Netherlands }
\affiliation {University of Maryland, College Park, MD 20742, USA }
\affiliation {Center for Relativistic Astrophysics and School of Physics, Georgia Institute of Technology, Atlanta, GA 30332, USA }
\affiliation {Laboratoire des Mat\'eriaux Avanc\'es (LMA), CNRS/IN2P3, F-69622 Villeurbanne, France }
\affiliation {Universit\'e Claude Bernard Lyon 1, F-69622 Villeurbanne, France }
\affiliation {Universit\`a di Napoli ``Federico II,'' Complesso Universitario di Monte S.Angelo, I-80126 Napoli, Italy }
\affiliation {NASA/Goddard Space Flight Center, Greenbelt, MD 20771, USA }
\affiliation {RESCEU, University of Tokyo, Tokyo, 113-0033, Japan. }
\affiliation {Tsinghua University, Beijing 100084, China }
\affiliation {Texas Tech University, Lubbock, TX 79409, USA }
\affiliation {The Pennsylvania State University, University Park, PA 16802, USA }
\affiliation {National Tsing Hua University, Hsinchu City, 30013 Taiwan, Republic of China }
\affiliation {Charles Sturt University, Wagga Wagga, New South Wales 2678, Australia }
\affiliation {West Virginia University, Morgantown, WV 26506, USA }
\affiliation {University of Chicago, Chicago, IL 60637, USA }
\affiliation {Caltech CaRT, Pasadena, CA 91125, USA }
\affiliation {Korea Institute of Science and Technology Information, Daejeon 305-806, Korea }
\affiliation {Universit\`a di Roma ``La Sapienza,'' I-00185 Roma, Italy }
\affiliation {University of Brussels, Brussels 1050, Belgium }
\affiliation {Sonoma State University, Rohnert Park, CA 94928, USA }
\affiliation {Center for Interdisciplinary Exploration \& Research in Astrophysics (CIERA), Northwestern University, Evanston, IL 60208, USA }
\affiliation {University of Minnesota, Minneapolis, MN 55455, USA }
\affiliation {The University of Melbourne, Parkville, Victoria 3010, Australia }
\affiliation {Institute for Plasma Research, Bhat, Gandhinagar 382428, India }
\affiliation {The University of Sheffield, Sheffield S10 2TN, United Kingdom }
\affiliation {The University of Texas Rio Grande Valley, Brownsville, TX 78520, USA }
\affiliation {Universit\`a di Trento, Dipartimento di Fisica, I-38123 Povo, Trento, Italy }
\affiliation {INFN, Trento Institute for Fundamental Physics and Applications, I-38123 Povo, Trento, Italy }
\affiliation {Cardiff University, Cardiff CF24 3AA, United Kingdom }
\affiliation {Montclair State University, Montclair, NJ 07043, USA }
\affiliation {MTA E\"otv\"os University, ``Lendulet'' Astrophysics Research Group, Budapest 1117, Hungary }
\affiliation {National Astronomical Observatory of Japan, 2-21-1 Osawa, Mitaka, Tokyo 181-8588, Japan }
\affiliation {School of Mathematics, University of Edinburgh, Edinburgh EH9 3FD, United Kingdom }
\affiliation {Indian Institute of Technology, Gandhinagar Ahmedabad Gujarat 382424, India }
\affiliation {University of Szeged, D\'om t\'er 9, Szeged 6720, Hungary }
\affiliation {Embry-Riddle Aeronautical University, Prescott, AZ 86301, USA }
\affiliation {Tata Institute of Fundamental Research, Mumbai 400005, India }
\affiliation {INAF, Osservatorio Astronomico di Capodimonte, I-80131, Napoli, Italy }
\affiliation {University of Michigan, Ann Arbor, MI 48109, USA }
\affiliation {Rochester Institute of Technology, Rochester, NY 14623, USA }
\affiliation {NCSA, University of Illinois at Urbana-Champaign, Urbana, Illinois 61801, USA }
\affiliation {Universitat de les Illes Balears, IAC3---IEEC, E-07122 Palma de Mallorca, Spain }
\affiliation {University of Bia{\l }ystok, 15-424 Bia{\l }ystok, Poland }
\affiliation {SUPA, University of Strathclyde, Glasgow G1 1XQ, United Kingdom }
\affiliation {IISER-TVM, CET Campus, Trivandrum Kerala 695016, India }
\affiliation {Canadian Institute for Theoretical Astrophysics, University of Toronto, Toronto, Ontario M5S 3H8, Canada }
\affiliation {Institute of Applied Physics, Nizhny Novgorod, 603950, Russia }
\affiliation {Pusan National University, Busan 609-735, Korea }
\affiliation {Hanyang University, Seoul 133-791, Korea }
\affiliation {University of Adelaide, Adelaide, South Australia 5005, Australia }
\affiliation {NCBJ, 05-400 \'Swierk-Otwock, Poland }
\affiliation {IM-PAN, 00-956 Warsaw, Poland }
\affiliation {Monash University, Victoria 3800, Australia }
\affiliation {Seoul National University, Seoul 151-742, Korea }
\affiliation {The Chinese University of Hong Kong, Shatin, NT, Hong Kong SAR, China }
\affiliation {University of Alabama in Huntsville, Huntsville, AL 35899, USA }
\affiliation {University of Massachusetts-Amherst, Amherst, MA 01003, USA }
\affiliation {ESPCI, CNRS, F-75005 Paris, France }
\affiliation {Universit\`a di Camerino, Dipartimento di Fisica, I-62032 Camerino, Italy }
\affiliation {Southern University and A\&M College, Baton Rouge, LA 70813, USA }
\affiliation {College of William and Mary, Williamsburg, VA 23187, USA }
\affiliation {Instituto de F\'\i sica Te\'orica, University Estadual Paulista/ICTP South American Institute for Fundamental Research, S\~ao Paulo SP 01140-070, Brazil }
\affiliation {University of Cambridge, Cambridge CB2 1TN, United Kingdom }
\affiliation {IISER-Kolkata, Mohanpur, West Bengal 741252, India }
\affiliation {Rutherford Appleton Laboratory, HSIC, Chilton, Didcot, Oxon OX11 0QX, United Kingdom }
\affiliation {Whitman College, 345 Boyer Avenue, Walla Walla, WA 99362 USA }
\affiliation {National Institute for Mathematical Sciences, Daejeon 305-390, Korea }
\affiliation {Universit\'e de Lyon, F-69361 Lyon, France }
\affiliation {Hobart and William Smith Colleges, Geneva, NY 14456, USA }
\affiliation {Janusz Gil Institute of Astronomy, University of Zielona G\'ora, 65-265 Zielona G\'ora, Poland }
\affiliation {King's College London, University of London, London WC2R 2LS, United Kingdom }
\affiliation {Andrews University, Berrien Springs, MI 49104, USA }
\affiliation {Universit\`a di Siena, I-53100 Siena, Italy }
\affiliation {Trinity University, San Antonio, TX 78212, USA }
\affiliation {University of Washington, Seattle, WA 98195, USA }
\affiliation {Kenyon College, Gambier, OH 43022, USA }
\affiliation {Abilene Christian University, Abilene, TX 79699, USA }



\maketitle

\acrodef{CW}{continuous wave}
\acrodef{CCO}{central compact object}
\acrodef{GW}{gravitational waves}
\acrodef{LHO}{LIGO Hanford Observatory}
\acrodef{LLO}{LIGO Livingston Observatory}
\acrodef{LIGO}{the Laser Interferometer Gravitational-wave Observatory}
\acrodef{LSC}{LIGO Scientific Collaboration}
\acrodef{PSD}{power spectral density}
\acrodef{S1}{the first initial LIGO science run}
\acrodef{S5}{the fifth initial LIGO science run}
\acrodef{S6}{the sixth initial LIGO science run}
\acrodef{SFT}{Short Fourier Transform}
\acrodef{SNR}{supernova remnant}
\acrodef{ULE}{upper limit estimate}
\acrodef{ULEs}{upper limit estimates}
\def\fscan{Fscan}

\section{Introduction}
The \ac{LSC} and Virgo Collaboration have undertaken numerous searches for continuous \ac{GW}. None has yet detected a signal, but many have placed interesting upper limits on possible sources. These searches have generally been drawn from one of three types.

Targeted searches are aimed at a single known pulsar, with a known precise timing solution.  The first search for continuous waves, using data from \ac{S1}, was of this type \cite{S1pulsar}, and subsequent searches have probed the Crab and Vela pulsars, among others \cite{S2pulsars, S3S4pulsars, S5Crab, VirgoVela, S5pulsars, S6pulsars}. A number of these most recent searches have been able to set direct upper limits on \ac{GW} emission comparable to or stricter than the indirect ``spin-down limits'' (derived from energy conservation, as well as the distance from Earth of the target, its gravitational-wave frequency, and the frequency's first derivative, the ``spin-down'') for a few of the pulsars searched.

All-sky searches, as their name suggests, survey the entire sky for neutron stars not seen as pulsars. These are very computationally costly, searching over wide frequency bands and covering large ranges of spin-down parameters \cite{S2Hough, S2Fstat, S4PSH, S4Einstein, S5PowerFlux, S5Einstein, S5PowerFlux2, S5Einstein2, S5Hough, VirgoAllSky}.  The latest of these have incorporated new techniques to cover possible binary parameters as well \cite{AllSkyBinary}. Recent all-sky searches have set direct upper limits close to indirect upper limits derived from galactic neutron-star population simulations \cite{KnispelAllen}.

Directed searches sit between these two extremes.  As in the all-sky case, their targets are neutron stars not seen as pulsars, so that the frequency and other parameters are unknown. They focus, however, on a known sky location (and therefore a known detector-frame Doppler modulation).  This directionality allows for searching over a wide range of frequencies and frequency derivatives while remaining much cheaper computationally than an all-sky search without sacrificing sensitivity. This approach was first used in a search for the accreting neutron star in the low-mass X-ray binary Sco~X-1 \cite{S2Fstat, S4radiometer, S5Stochastic}.

The search for the \ac{CCO} in the \ac{SNR} Cassiopeia A (Cas~A)\cite{S5CasA} was the first directed search for a young neutron star without electromagnetically detected pulsation, motivated by the idea that young neutron stars might be promising emitters of continuous \ac{GW}. The Cas~A search \cite{S5CasA} set upper limits on \ac{GW} strain which beat an indirect limit derived from energy conservation and the age of the remnant \cite{Wette2008} over a wide frequency band.  Other directed searches have since followed in its footsteps, using different data analysis methods, for supernova 1987A and unseen neutron stars near the galactic center \cite{S5Stochastic, S5GalacticCenter}. Most methodologically similar to this search and the S5 Cas~A search was a recent search for nine supernova remnants \cite{S6CasAFriends}, which also used fully coherent integration over observation times on the order of 10 days.

In this article, we describe a search of data from \ac{S6} for potential young isolated neutron stars with no observed electromagnetic pulsations in the nearby ($d \approx 2.7$ kpc) globular cluster NGC~6544. Globular clusters are unlikely to contain young neutron stars, but in these dense environments older neutron stars may be subject to debris accretion (see Sec.~\ref{s:TS}) or other events which could render them detectable as gravitational wave sources. This particular globular cluster was chosen so that a computationally-feasible coherent search similar to \cite{S5CasA} could beat the age-based indirect limits on \ac{GW} emission. 

The search did not find a \ac{GW} signal, and hence the main result is a set of upper limits on strain amplitude, fiducial ellipticity, and $r$-mode amplitude $\alpha $, similar to those presented in \cite{S5CasA}. An important new feature of the search described here was use of a barycentric resampling algorithm which substantially reduced computational cost, allowing a search over a larger parameter space using a longer coherence time (see Sec.~\ref{s:SPS}). This barycentric resampling method will be used extensively in searches of Advanced LIGO and Virgo detector data. 

This article is structured as follows: In Sec.~\ref{s:searches} we present the method, implementation, and results of the search. The upper limits set in the absence of a signal are presented in Sec.~\ref{s:uls}, and the results are discussed in Sec.~\ref{s:disc}.

\section{Searches}
\label{s:searches}

\subsection{Data selection}
The sixth initial LIGO science run (S6) extended from July 7 2009 21:00:00 UTC (GPS 931035615) to October 21 2010 00:00:00 UTC (GPS 971654415) and included two initial LIGO detectors with 4-km arm lengths, H1 at \ac{LHO} near Hanford, Washington and L1 at \ac{LLO} near Livingston, Louisiana.

After optimization at fixed computing cost determined an optimum coherence time of 9.2 days (see Sec.~\ref{s:SPS}), two different methods were used to determine which data would be searched, producing two different 9.2-day stretches. Both were searched, allowing for the comparison of search results between them.

The first method was to look for the most sensitive average data from S6. This was done by taking nine week-long data samples from each detector spaced roughly 55 days apart, giving nine evenly spaced weeks throughout the duration of S6. The data samples used are shown in Table ~\ref{SampleTable}.  We chose four representative frequencies ($100$ Hz, $200$ Hz, $400$ Hz, $600$ Hz) and generated joint-detector strain noise power spectral densities (PSDs) in $1$-Hz bands about these frequencies, using $0.01$-Hz binning. The sensitivity $h_{\textup{sens}}$ was then taken to be

\begin{equation}
\label{SensitivityEquation}
h_{\textup{sens}}^{j} = \left ( \frac{1}{\sqrt{(1/100) \cdot \sum_{i = 0}^{100}(S_{h}^{i}(f_{i}))^{-1}}} \right )_{j}
\end{equation}

\noindent where $S_{h}^{i}(f)$ represents the PSD value of the $i^{th}$ bin, at frequency $f_{i}$, and the index $j$ runs from 1 through 4 and represents the four representative frequencies (note that this is not an actual estimate of detectable strain). Based on this figure of merit, the final nine days of S6 yielded the most sensitive data stretch for all four frequencies: October 11-20, 2010 (GPS 970840605 -- 971621841).

\begin{table} \centering \begin{tabular} {| c | c | c | c |}\hline \multicolumn{4}{|c|}{S6 sampling times} \\ \hline Label & GPS Start & GPS End & Dates (UTC) \\ \hline Week 1 & 931053000 & 931657800 & Jul 8-15, 2009 \\ \hline Week 2 & 936053000 & 936657800 & Sep 3-10, 2009 \\ \hline Week 3 & 941053000 & 941657800 & Oct 31-Nov 7, 2009 \\ \hline Week 4 & 946053000 & 946657800 & Dec 28, 2009-Jan 4, 2010 \\ \hline Week 5 & 951053000 & 951657800 & Feb 24-Mar 3, 2010 \\ \hline Week 6 & 956053000 & 956657800 & Apr 23-30, 2010 \\ \hline Week 7 & 961053000 & 961657800 & Jun 20-27, 2010 \\ \hline Week 8 & 966053000 & 966657800 & Aug 17-24, 2010 \\ \hline Week 9 & 971053000 & 971657800 & Oct 14-21, 2010 \\ \hline \end{tabular}
\caption{The weeks sampled to find the most sensitive S6 data. Times are given both in GPS and UTC calendar dates.}
\label{SampleTable}
\end{table}

An alternate data selection scheme \cite{S6CasAFriends, S5CasA}, which takes detector duty cycle into account is to maximize the figure of merit

\begin{equation}
\label{FOM}
\sum_{k,f} \frac{1}{S_h(f)}
\end{equation}

\noindent where $S_h(f)$ represents the strain noise power spectral density at frequency $f$ in the $k^{th}$ \ac{SFT}, and the sum is taken across all frequencies $f$ in the search band and all SFTs in a given 9.2-day (see Sec.~\ref{s:SPS} below) observation time. The \ac{SFT} format is science-mode detector data split into 1800s segments, band-pass filtered from 40--2035~Hz, Tukey windowed in the time domain, and Fourier transformed. This method favored a different data stretch: July 24--August 2, 2010 (GPS 964007133 -- 964803598). This second data stretch had slightly worse average sensitivity than the first, but a higher detector livetime: our first (October) data set contained 374 SFTs (202 from Hanford and 172 from Livingston) with average sensitivity $h_{sens}^{200\textup{Hz}} = 1.92 \times 10^{-23}$; the second (July-August) data set contained 642 SFTs (368 from Hanford and 274 from Livingston) with average sensitivity $h_{sens}^{200\textup{Hz}} = 1.95 \times 10^{-23}$.

\subsection{Analysis method}
\label{s:A_method}
The analysis was based on matched filtering, the optimal method for detecting signals of known functional form. To obtain that form we assumed that the potential target neutron star did not glitch (suffer an abrupt frequency jump) or have significant timing noise (additional, possibly stochastic, time dependence of the frequency) \cite{LyneSmith2006} during the observation. We also neglected third and higher derivatives of the \ac{GW} frequency, based on the time span and range of $\dot{f}$ and $\ddot{f}$ (the first two derivatives) covered. The precise expression for the interferometer strain response $h(t)$ to an incoming continuous \ac{GW} also includes amplitude and phase modulation by the changing of the beam patterns as the interferometer rotates with the earth. It depends on the source's sky location and orientation angles, as well as on the parameters of the interferometer. The full expression can be found in \cite{Jaranowski1998}.

The detection statistic used was the multi-interferometer $\mathcal{F}$-statistic \cite{Cutler2005}, based on the single-interferometer $\mathcal{F}$-statistic \cite{Jaranowski1998}. This combines the results of matched filters for the signal in a way that is computationally fast and nearly optimal \cite{Prix2009}. Assuming Gaussian noise, $2\mathcal{F}$ is drawn from a $\chi^2$ distribution with four degrees of freedom.

We used the implementation of the $\mathcal{F}$-statistic in the LALSuite package \cite{lalsuite}. In particular most of the computing power of the search was spent in the \texttt{ComputeFStatistic\_v2} program. Unlike the version used in preceding methodologically similar searches \cite{S5CasA,S6CasAFriends}, this one implements an option to use a barycentric resampling algorithm which significantly speeds up the analysis.

The method of efficiently computing the $\mathcal{F}$-statistic by barycentering and Fast-Fourier-transforming the data was first proposed in \cite{Jaranowski1998}. Various implementations of this method have been developed and used in previous searches, such as \cite{Krolak2010,PinkeshMethods,PatelThesis}. Here we are using a new LALSuite \cite{lalsuite} implementation of this method, which evolved out of \cite{PatelThesis}, and which will be described in more detail in a future publication. It converts the input data into a heterodyned, downsampled timeseries weighted by antenna-pattern coefficients, and then resamples this timeseries at the solar system barycenter using an interpolation technique.  The resampled time series is then Fourier-transformed to return to the frequency domain, and from there the $\mathcal{F}$-statistic is calculated. For this search, both single-detector and multi-detector $\mathcal{F}$-statistics were calculated (see Vetoes section below).

Timing tests run on a modern processor (ca. 2011) showed that the resampling code was more than 24 times faster in terms of seconds per template per SFT. This improvement, by more than an order of magnitude, was used to perform a deeper search over a wider parameter space than previously possible for the computational cost incurred (see target selection and search parameters below).

\subsection{Target selection}
\label{s:TS}
Unlike previous directed searches, this one targets a globular cluster. Since stars in globular clusters are very old, it is unlikely that a young neutron star will be found in such an environment. However, some neutron stars are known to be accompanied by debris disks \cite{Wang} and even planets \cite{WolszczanFrail, Thorsett, Sigurdsson2003}. In the densely populated core of a globular cluster, close encounters may stimulate bombardment episodes as debris orbits are destabilized, akin to cometary bombardments in our solar system when the Oort cloud is perturbed \cite{Sigurdsson1992}. A neutron star which has recently accreted debris could have it funneled by the magnetic field into mountains which relax on timescales of $10^{5}$--$10^{8}$ years \cite{Vigelius} and emit gravitational waves for that duration. Other mechanisms are likely to last a few years at most \cite{Haskell}. Hence an old neutron star could be a good gravitational wave source with a low spin-down age.

The first step in picking a globular cluster is a figure of merit based on that for directed searches for supernova remnants \cite{Wette2008}, an indirect upper limit on gravitational wave strain based on energy conservation and the age of the object. Here the inverse of the object age is replaced by the interaction rate of the globular cluster, which scales like density$^{(3/2)}$ times core radius$^{2}$ \cite{VerbuntHut, Sigurdsson1992}, reflecting the mean time since last bombardment. It is hard to know when the most recent bombardment episode was, and thus the constant factor out in front, but globular clusters can be ranked with respect to each other by a maximum-strain type figure of merit

\begin{equation}
\label{GCFOM}
h_{0} \propto \rho_{c}^{3/4} r_{c}/d,
\end{equation}

\noindent where $\rho_{c}$ is the globular cluster core density, $r_{c}$ is the core radius, $d$ is the distance to the cluster, and thus $r_{c}/d$ is the angular radius of the core. We ranked the Harris catalog of globular clusters \cite{Harris1, Harris2} by this figure of merit and looked at the top few choices, which were mainly nearby core-collapsed clusters. The closest is NGC~6397 at $\approx 2.2$ kpc, but it is at high declination. This lessens the Doppler modulation of any gravitational wave signal, making it harder to distinguish from stationary spectral line artifacts, which tend to contaminate searches at high declination near the ecliptic pole.  Hence we chose the next closest, NGC~6544, which is at a declination of less than 30 degrees and only slightly further away at $\approx 2.7$ kpc.

We restrict the search described below to sources for which the bombardment history corresponds to a characteristic spindown age older than 300 years. The figure of 300 years is mainly a practical consideration: the cost of a search rises steeply for lower spin-down ages, and 300 years proved tractable for the Cas~A search \cite{S5CasA}.

\subsection{Search parameter space}
\label{s:SPS}
An iterative method was used to generate the parameter space to be searched.  Starting with an (assumed) spin-down age no younger than 300 years, a braking index $n$ = 5 (see below), and the known distance to the globular cluster, we calculated the age-based indirect upper limit. This is an optimistic limit on the gravitational wave strain $h_0$ which assumes that all energy lost as the target neutron star spins down is radiated away as gravitational waves\cite{Wette2008}:

\begin{equation}
\label{indirect1}
h_0 \leq \frac{1}{d} \sqrt{\frac{5 G I}{2 {c}^{3}\tau (n-1)}}.
\end{equation}

\noindent Here $d$ is the distance to the target, $\tau $ the assumed age of the target object, and $I$ a fiducial moment of inertia for a neutron star ($10^{38} \textup{kg} \cdot \textup{m}^{2}$). $G$ and $c$ are the gravitational constant and the speed of light, respectively.  This age-based limit was then superimposed on a curve of expected upper limits in the absence of signal for the LIGO detectors, obtained from the noise \ac{PSD} harmonically averaged over all of \ac{S6} and both interferometers. A running median with a 16-Hz window was further applied to smooth the curve. The curve is given by:

\begin{equation}
\label{senscurve}
h_{0}^{95\%} = \Theta \sqrt{ \frac{S_h} {T_\mathrm{data}} }
\end{equation}

\noindent where $S_{h}$ is the harmonically averaged noise, $T_\mathrm{data}$ is the coherence time (the total data livetime searched coherently), initially estimated at two weeks, and $\Theta$ is a sensitivity factor that includes a trials factor, or number of templates searched, and uncertainty in the source orientation \cite{Wette2008}. For a directed search like ours, $\Theta$ is approximately 35 \cite{Wette2008, WetteSens}.  The intersection of this coherence-time adjusted upper limit curve and our indirect limit (Eq.~(\ref{indirect1})) gives an initial frequency band over which the indirect limit can be beaten. The braking index is related to the frequency parameters by the definition:

\begin{equation}
\label{braking_index}
n = \frac{f\ddot{f}}{\dot{f}^{2}}.
\end{equation}

\noindent Assuming a braking index $n$ between 2 and 7 covers most accepted neutron star models ($n=5$, the neutron star radiating all energy as gravitational waves via the mass quadrupole, is used to obtain the indirect limit). We allow the braking indices in these expressions to range from 2 to 7 independently, to reflect the fact that in general multiple processes are operating and $\dot{f}$ is not a simple power law. This constraint on the braking indices then produces limits on the frequency derivatives given by \cite{Wette2008}

\begin{equation}
\label{spindown range}
\frac{f} {\tau } \leq -\dot{f} \leq \frac{f} {6\tau}
\end{equation}

\noindent for the spindown at each frequency and

\begin{equation}
\label{second spindown range}
\frac{2 \dot{f}^2} {f} \leq \ddot{f} \leq \frac{7 \dot{f}^2} {f}
\end{equation}

\noindent for the second spindown at each $(f,\dot{f})$. The step sizes for frequency and its derivatives are given by the equations \cite{WhitbeckThesis2006,PatelThesis, ReinhardStep}

\begin{equation}
\label{freqstep}
df = \frac{2\sqrt{3m}}{\pi }\frac{1}{T_\mathrm{data}},
\end{equation}

\begin{equation}
\label{fdotstep}
d\dot{f} = \frac{12\sqrt{5m}}{\pi }\frac{1}{T^{2}_\mathrm{data}},
\end{equation}

\noindent and

\begin{equation}
\label{fddotstep}
d\ddot{f} = \frac{20\sqrt{7m}}{\pi }\frac{1}{T^{3}_\mathrm{data}}.
\end{equation}  

\noindent where $m$ is the mismatch parameter, the maximum loss of $2\mathcal{F}$ due to discretization of the frequency and derivatives \cite{Owen1996, Brady1998}. This search used a mismatch parameter $m = 0.2$.

From these relations the total number of templates (points in frequency parameter space) to be searched can be calculated, and with knowledge of the per-template time taken by the code (obtained from timing tests),  the total computing time can be obtained. Limiting the target computing time, in our case to 1000 core-months, then allows us to solve for the coherence time $T_\mathrm{data}$, which we then feed back into Eq.~(\ref{senscurve}) to begin the process anew until it iteratively converges on a parameter space and accompanying coherence time. The iterative algorithm thus balances the computational gains from resampling between the use of a longer coherence time (giving better sensitivity) and the expansion of the parameter space over which the indirect limit can be beaten (caused by the improved sensitivity). The result for the globular cluster NGC~6544 is a search over the frequency range 92.5 Hz to 675 Hz, with a coherence time of 9.2 days.

The peculiar velocities of globular clusters are negligible, as they represent an essentially constant Doppler shift of order $1 \times 10^{-3}$; so is velocity dispersion, which is an order of magnitude smaller. Since we search down to 300-year timescales, the acceleration of the cluster is also not an issue \cite{MillionBody}.

\subsection{Implementation}
All searches were run on the LIGO-Caltech Computing Cluster at the California Insitute of Technology in Pasadena, CA, under the control of the Condor queuing system for parallel processing. The search process was split into 5825 individual Condor jobs, each of which searched over a $0.1$-Hz subband and corresponding swathes of $(\dot{f},\ddot{f})$. The number of templates searched by each job thus varied as a function of frequency. 

Each search job produced three distinct outputs. First, a record was made of all candidates with $2\mathcal{F}$ above 45.0, a choice of recording different from the \ac{S5} search which recorded the loudest $0.01\%$ of events. This was needed because of the contamination of the \ac{S6} noise by detector artifacts, as well as limits on the disk space available and the input/output capability of the cluster filesystem. Second, a histogram of $2\mathcal{F}$ values for all templates searched was produced to verify that the data matched the expected chi-square distribution (described in Subsec.~\ref{s:A_method} above). Last, each job produced a record of the loudest (highest-$2\mathcal{F}$-valued) candidate in its 0.1-Hz band, regardless of threshold. This data was used in the setting and validation of upper limits (see Section ~\ref{s:uls} below).

\subsection{Vetoes}
\begin{table}
\begin{center}
\resizebox{0.47\textwidth}{!}{
\begin{tabular}{lrrl}
\hline
\hline
Band & \multicolumn{2}{c}{Job min.\ and max.} & Note
\\
& \multicolumn{2}{c}{frequency (Hz)} &
\\
\hline
370.1 & 370.1 & 370.2 & L1 Output Mode Cleaner (OMC) Jitter Line
\\
393.1 & 393.1 & 393.2 & H1 Calibration Line
\\
396.7 & 396.7 & 396.8 & L1 Calibration Line
\\
400.2 & 400.2 & 400.3 & H1 OMC Quad Photodiode (QPD) Line
\\
403.8 & 403.8 & 403.9 & L1 OMC QPD Line
\\
417.1 & 417.1 & 417.2 & H1 OMC QPD Line
\\
580.0 & 580.0 & 580.1 & L1 2Hz Harmonic
\\ 
\hline
\end{tabular}}
\end{center}
\caption{Search sub-bands that, due to the identified disturbances, produced an excessive number of candidates and were aborted. The 580.0 Hz sub-band had to be stopped only for the July-August run; the other six bands were vetoed in both searches.}
\label{t:BadBands}
\end{table}

A high value of $2\mathcal{F}$ is not enough to claim a detection, since instrumental artifacts lead to non-Gaussian and/or non-stationary noise in many narrow frequency bands. A variety of veto techniques were used to trim down the initial list of candidates and arrive at a final list of outliers.

Six 0.1 Hz sub-bands (see Table~(\ref{t:BadBands})) had to be manually aborted in both searches, with a seventh aborted in the July-August search, as even with the threshold in place, they produced an excessive number of candidates. Each of these subbands was compared to records of known noise artifacts and disturbances in the detector, and in each case a known instrumental line was confirmed. These sub-bands were later rerun with the record of candidates disabled in order to produce histograms and loudest-outlier files for upper limit validation.

To protect against spurious noise lines, a second veto based on the $\mathcal{F}$-statistic consistency veto introduced in \cite{S5Einstein2} was used. This uses the fact that an astrophysical signal should have a higher joint value of $2\mathcal{F}$ (combining data from the two interferometers) than in either interferometer alone. Recorded candidates that violate this inequality were vetoed. This is a simpler version of the more recent line veto \cite{Keitel2014}.

Finally, to enforce coincidence between detectors, a single-detector threshold was employed.  Since a true astrophysical signal should be present in both detectors at a significant level, any candidates passing the initial joint-detector detection criteria (see ~\ref{s:DCR}) also had to pass an additional threshold on the individual-detector values of $2\mathcal{F}$.

The 0.1-Hz band between 200 and 200.1 Hz was arbitrarily chosen as a test band. The joint-detector $2\mathcal{F}$ values were taken from the loudest-candidate files and used to semi-analytically compute ~\cite{Wette2009} an estimate of the 95\% upper limit for that subband using the SFTs employed by the search. Sets of 1,000 software injections were performed with strengths of 100\%, 80\%, 60\%, 40\% and 20\% of this estimated upper limit. The results were used to set a threshold of $2\mathcal{F} \geq 20$ in each individual detector, leading to an additional false dismissal rate of 1.5\% of injections at the 95\% confidence upper limit estimate. Candidates failing to meet this criterion were vetoed.

\subsection{Detection criteria and results}
\label{s:DCR}

The results of a mock data challenge were used to set a detection criterion for the joint-detector $2\mathcal{F}$ value. The mock data challenge consisted of a set of 1577 artificial \ac{CW} signals injected into a set of real detector data from S6, which were then searched for using the same resampled $\mathcal{F}$-Statistic used in the search. A survey of the loudest joint-detector $2\mathcal{F}$ value reported for background subbands known to be free of injected signals for the band between 200 Hz and 240 Hz (used in a pilot run) gave a mean loudest joint-detector $2\mathcal{F} \approx 55$. Given this background level, the detection criterion was chosen to be joint-detector $2\mathcal{F} \geq 60$ to maintain high efficiency and low false-alarm rate (the false-alarm rate was 3.17\% in these pilot subbands).

With these detection criteria, a search was carried out in \ac{S6} data. The lists of all templates with joint-detector $2\mathcal{F}$ greater than 45.0 were filtered for the individual detector threshold and the consistency veto, both singly and in tandem. If the loudest template failed either check, the list was used to move to the next-loudest template until the loudest template passing all thresholds and vetoes was identified. This created three sets of results (threshold-only, veto-only, and threshold+veto) which could all be queried independently. 

The joint $2\mathcal{F}$ values for the loudest single template (passing all thresholds and vetoes) in each 0.1 Hz subband were collated into lists spanning 10Hz (100 joint $2\mathcal{F}$ values per list). These lists were then parsed, and any joint $2\mathcal{F}$ values greater than the joint $2\mathcal{F}$ threshold of 60 were identified. Each such entry's corresponding template was then added to a list of outliers. This method produced a list of 168 outliers for the entirety of the search band in the October data, and a list of 155 outliers for the entirety of the search band in the July-August data.

These outliers were then tested using \emph{time shifts} and \emph{extended looks}. In a time shift, the frequency parameters of the outliers from each data stretch (October and July-August) were evolved forwards or backwards in time, as appropriate, and sought in the opposite data stretch, under the assumption that a true astrophysical signal should be present in both data sets for the implicitly long-lived \ac{CW} signals searched for here. A set of 1000 software injections (simulated signals with randomly generated parameters) underwent the same treatment to provide a baseline $2\mathcal{F}$ threshold for signal detection, and outliers surpassing the threshold were considered present.

In an extended look, each outlier was sought in an expanded 20-day coherence time encompassing the original nine-day coherence time; the same assumption of signal continuity would predict, roughly, a doubling of the $2\mathcal{F}$ value for a doubling of coherence time.  These cases as well were tested with software injections to determine a threshold. 

In both time shifts and extended looks, the searches were conducted over a parameter space envelope obtained by starting at the outlier frequency parameters $(f, \dot{f}, \ddot{f}) \pm 2$ bins, and evolving those ranges backwards or forwards in time using the extremum values of the next derivative (e.g., $f$ evolved at maximum $\dot{f}$, $\dot{f}$ evolved at maximum $\ddot{f}$) to achieve a conservatively wide envelope.

Outliers detected with joint $2\mathcal{F}$ greater than the threshold established by the software injections were labeled candidates and received manual followup. These tests were not cumulative; an outlier needed only to survive any one test, not all of them, to persist as a candidate.  The software injection threshold for both types of test was placed at a value for joint $2\mathcal{F}$ yielding 80\% injection recovery; because each outlier would receive further consideration if it passed either test, the false dismissal probability for the first follow-up stage was $\approx 4\%$. The combined 323 outliers produced only seven candidates, listed in Table ~\ref{CandidateTable}.

\begin{table*}[t]
\begin{center}
\resizebox{0.98\textwidth}{!}{
\begin{tabular} {| c | c | c | c | c | p{3cm} |}\hline \multicolumn{6}{|c|}{July-August Data} \\ \hline Outlier & Search $f$ (Hz) & Followup $f$ (Hz) & Search $2\mathcal{F}_{J}$ & Followup $2\mathcal{F}_{J}$ & Artifact, if any\\ \hline 27 & 192.4907 & 192.4956 & 612.969 & 300.712 & Hardware Injection\\ \hline 74 & 392.2232 & 392.2315 & 189.903 & 173.787 & Clock noise\\ \hline 77 & 394.0231 & 394.0307 & 228.268 & 197.300 & Digital line\\ \hline \multicolumn{6}{|c|}{October data} \\ \hline 27 & 192.4195 & 192.4313 & 875.575 & 484.254 & Hardware Injection\\ \hline 79 & 403.6424 & 403.8612 & 114.626 & 61.331 & -----\\ \hline 85 & 417.0394 & 417.1384 & 60.309 & 176.200 & H1 Output Mode Cleaner Line\\ \hline 131 & 575.9658 & 576.5057 & 61.943 & 53.805 & -----\\ \hline \end{tabular}}
\caption{The seven candidates that passed the first round of outlier followup. The columns give, respectively: the outlier's identifying number; the frequency of the outlier in the search; the frequency of the outlier in the followup data set in which it appeared; the $2\mathcal{F}$ value of the outlier in the search; the $2\mathcal{F}$ value of the outlier in the followup data set in which it appeared; the explanation, if any, provided by comparison with run-averaged strain histograms in conjunction with detector characterization records.}\label{CandidateTable}
\end{center}
\end{table*}

These candidates were subject to manual followup. They were compared to strain histograms of run-averaged (i.e., over all of S6) spectra from each detector, to identify instrumental noise lines which could be responsible.  In five of the seven cases, the strain histograms gave clear evidence of an instrumental noise line responsible for the candidate, and in these cases records of prior detector characterization studies were consulted to provide explanations for the noise artifacts. In those cases the artifact is listed in Table ~\ref{CandidateTable} as well. Two of the artifacts arose from hardware injections at different points in the sky, used to test interferometer response ~\cite{S6CasAFriends}.

The remaining two candidates were given another round of followup, with a time shift and extended look performed in data from June 2010, the farthest removed (in the time domain) available data of comparable sensitivity. The large time separation creates a large difference in the Doppler corrections needed to reconstruct an astrophysical source, making these corrections unlikely to reinforce instrumental or environmental artifacts. Both outliers failed to pass the $2\mathcal{F}$ thresholds established by software injections in any of their June tests.

The loudest $2\mathcal{F}$ value expected in the absence of signal depends on the number of templates searched~\cite{Wette2009};\footnote{The $N_{T}$ templates used in our searches are not completely independent, but can be represented by $N$ statistically independent templates where $N \approx 0.88N_{T}$. See section 8.7 of ~\cite{Wette2009}.} for our search, the largest expected $2\mathcal{F}$ value lies in the range $72 \leq 2\mathcal{F} \leq 80$ with 90\% confidence. The $2\mathcal{F}$ values associated with the two remaining candidates, outliers 79 and 131, were joint-detector $2\mathcal{F} = 61.3$ and joint-detector $2\mathcal{F} = 61.9$, respectively. The outliers' failure to pass the June tests and their marginal $2\mathcal{F}$ values led us to dismiss them as noise fluctuations.

Thus no credible gravitational wave signals were detected by our search. In the absence of a detection, we can set upper limits on the possible strength of gravitational waves in our data.

\section{Upper limits}
\label{s:uls}

\subsection{Methods}
\begin{figure}
  \includegraphics[width=0.49\textwidth]{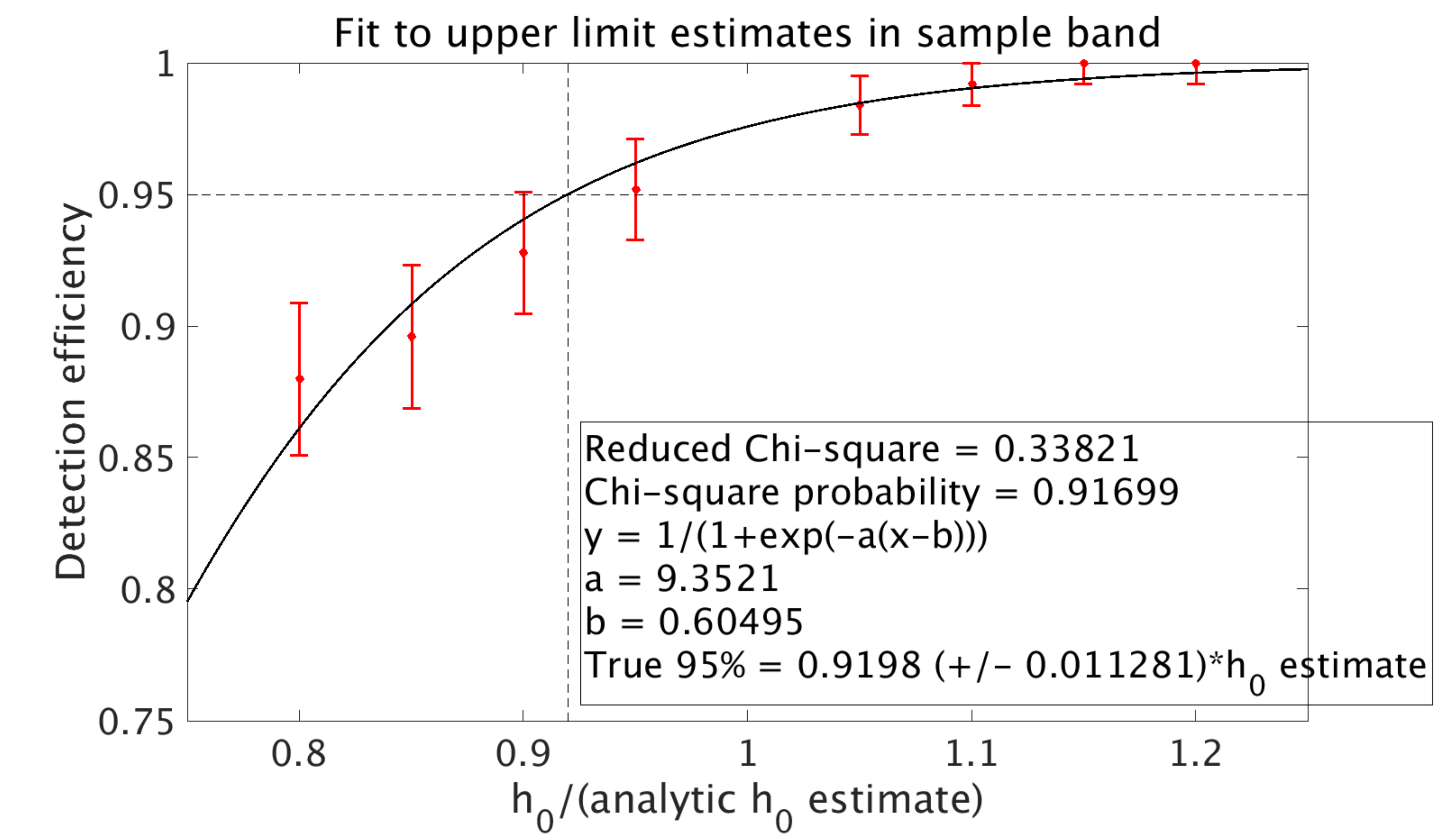}
  \caption{A demonstration of the upper limit validation technique for a sample band (101 Hz; October data). The x-axis is $h_{nom}$, $h_0$ divided by the \ac{ULE} for this upper-limit band; the eight points represent detection efficiencies for eight sets of 125 software injections.  These eight points are then fit to a sigmoid curve (in black); the 95\% upper limit can then be read off from the point where the curve crosses 95\% detection efficiency.}
  \label{ExamplePlot}
\end{figure}

The method of setting upper limits was a variation on that used in \cite{S5CasA}and \cite{S6CasAFriends}. This upper limit determination is based only of the $\mathcal{F}$-statistic and does not include additional criteria involved in candidate followup. We split the frequency band into 0.1-Hz subbands, and for each of these used a semi-analytic Monte Carlo method to estimate a 95\% upper limit, defined as the strain $h_0$ at which our detection criterion would successfully detect 95\% of signals.  Due to the high computational cost of individually verifying all 5800 such subbands, these 0.1-Hz upper limit bands were consolidated into 1-Hz subbands.  For each such 1-Hz band, we performed 1,000 software injections, split into eight groups of 125 signals. The strain $h_0$ of each group was $\pm 5\%, \pm 10\%, \pm 15\%, \textup{and} \pm 20\%$ of the semi-analytic \ac{ULEs}, respectively.  A software injection was considered validated if it returned a value of $2\mathcal{F}$ greater than or equal to the loudest outlier in its 0.1-Hz subband, thus maintaining the original granularity.  For each 1-Hz band, these 1,000 injections thus produced eight points on a detection efficiency curve.  We then used a least-squares method to produce a sigmoid fit to the data points, and from this curve determined a true 95\% upper limit, defined as the value of $h_0$ at which the fitting curve intersected 95\% efficiency. Figure ~\ref{ExamplePlot} shows such a plot for a sample band. In cases where the 95\% point was extrapolated (as opposed to interpolated) from the eight points, and in cases where the uncertainty on the 95\% point was greater than 5\%, a new set of eight points were generated using the 95\% point as the initial $h_{0}$ estimate and a 95\% point was determined from the combined sets (e.g., a curve was fit to 16 points after one rerun, 24 points after two reruns, etc.).

A small number of $0.1$-Hz bands had outliers so large that the semi-analytic method failed to converge to an estimate for $h_{0}$. Instrumental artifacts at these frequencies were identified using \ac{S6} run-averaged spectra and the respective 1-Hz bands were then rerun with the disturbed $0.1$-Hz subband excluded. The excluded bands are detailed in Table ~\ref{ExclBand}. 

\begin{table}
\begin{center}
\resizebox{0.49\textwidth}{!}{
\begin{tabular} {| c | c | p{3cm} |}\hline Affected 1-Hz Band & Vetoed 0.1-Hz subband & Artifact\\ \hline 180 & 180.0-180.1 & Power mains harmonic\\ \hline 192 & 192.4-192.5 & Hardware injection\\ \hline 217 & 217.5-217.6 & Known instrumental artifact\\ \hline 234 & 234.0-234.1 & Digital line (L1)\\ \hline 290 & 290.0-290.1 & Digital line (L1)\\ \hline 370 & 370.1-370.2 & L1 Output Mode Cleaner Line\\ \hline 393 & 393.1-393.2 & Calibration line (H1)\\ \hline 396 & 396.7-396.8 & Calibration Line (L1)\\ \hline 400 & 400.2-400.3 & H1 Output Mode Cleaner Line\\ \hline 403 & 403.8-403.9 & L1 Output Mode Cleaner Line\\ \hline 417 & 417.1-417.2 & H1 Output Mode Cleaner Line\\ \hline 580 & 580.0-580.1 & Digital line (L1)\\ \hline \end{tabular}}
\caption{The twelve $0.1$-Hz bands with outliers too large for the semi-analytic method to converge to an estimate for $h_{0}$. These $0.1$-Hz subbands were excluded from upper limit analysis; the quoted upper limits (Figure ~\ref{ULE}) represent the remainder of their respective 1-Hz bands. The first column lists the affected 1-Hz band; the second column lists the respective vetoed 0.1-Hz subband; the third column lists the instrumental artifact identified using \ac{S6} run-averaged spectra.}
\label{ExclBand}
\end{center}
\end{table}

\subsection{Results}

\begin{figure*}[htbp]
  \includegraphics[width=0.98\textwidth]{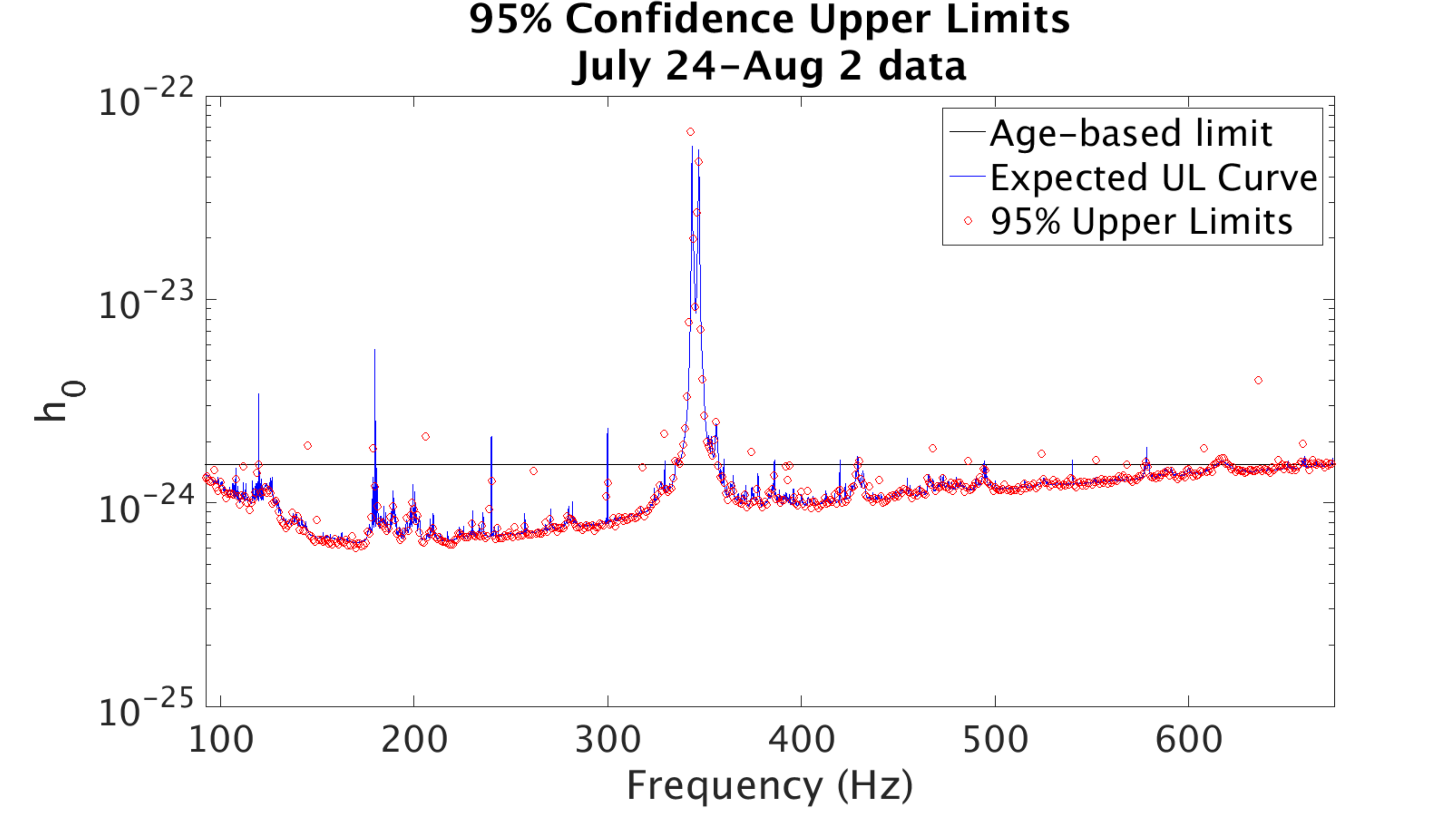}
  \caption{Upper limits at 95\% confidence (red circles) compared to the upper limit estimate curve of the detector (blue curve) and the initial age-based limit on $h_{0}$ (black line). The upper limit estimate curve is based on Eq.~(\ref{senscurve}) over the 9.2 days of coherence time, corrected for detector livetime and sky location and summed over detectors in inverse quadrature.}
  \label{ULE}
\end{figure*}

\begin{figure}[htbp]
  \includegraphics[width=0.49\textwidth]{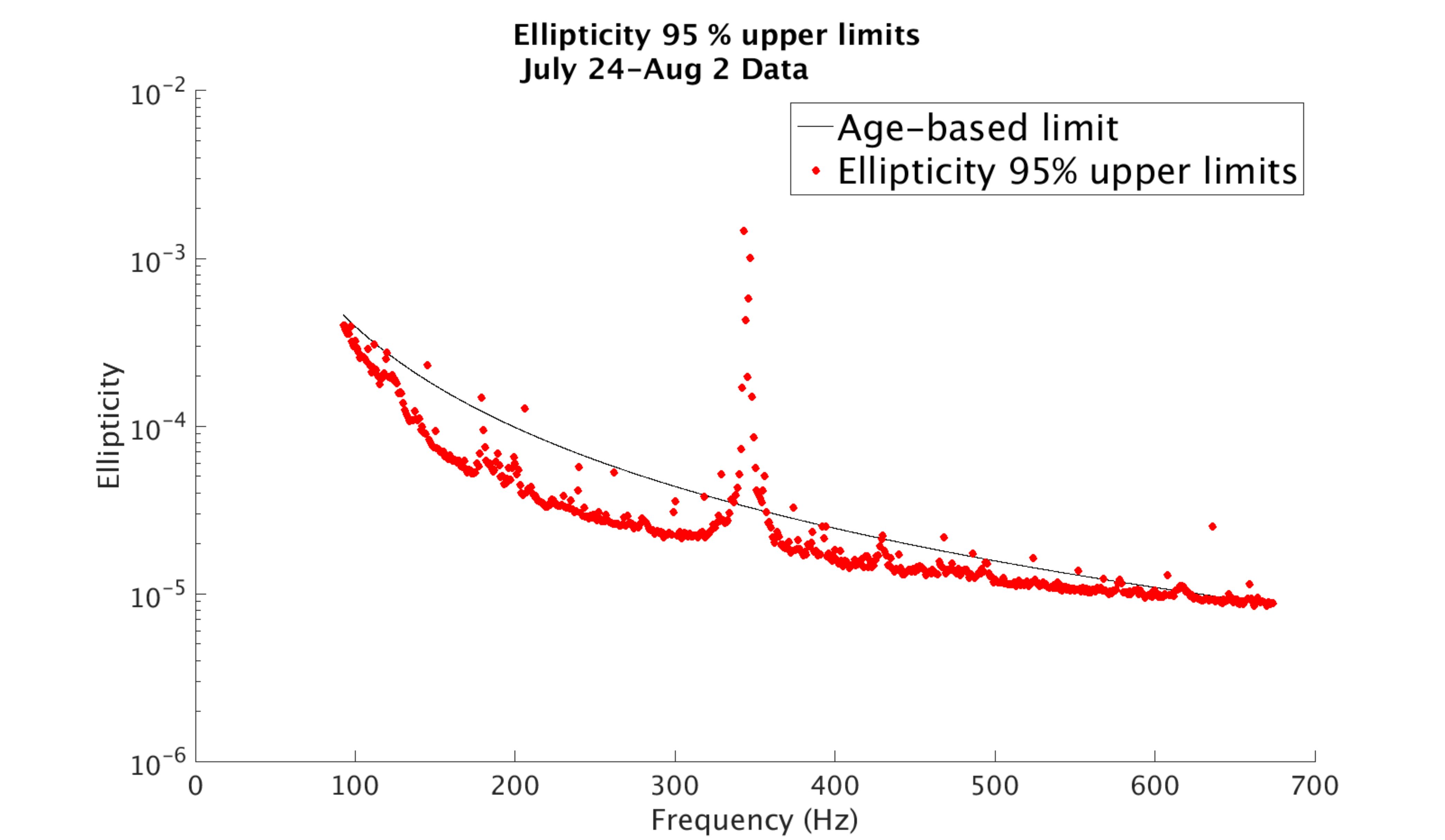}
  \caption{Upper limits at 95\% confidence (red circles) compared to the initial age-based limit on fiducial ellipticity $\epsilon $ (black line).}
  \label{Ellip}
\end{figure}

\begin{figure}[htbp]
  \includegraphics[width=0.49\textwidth]{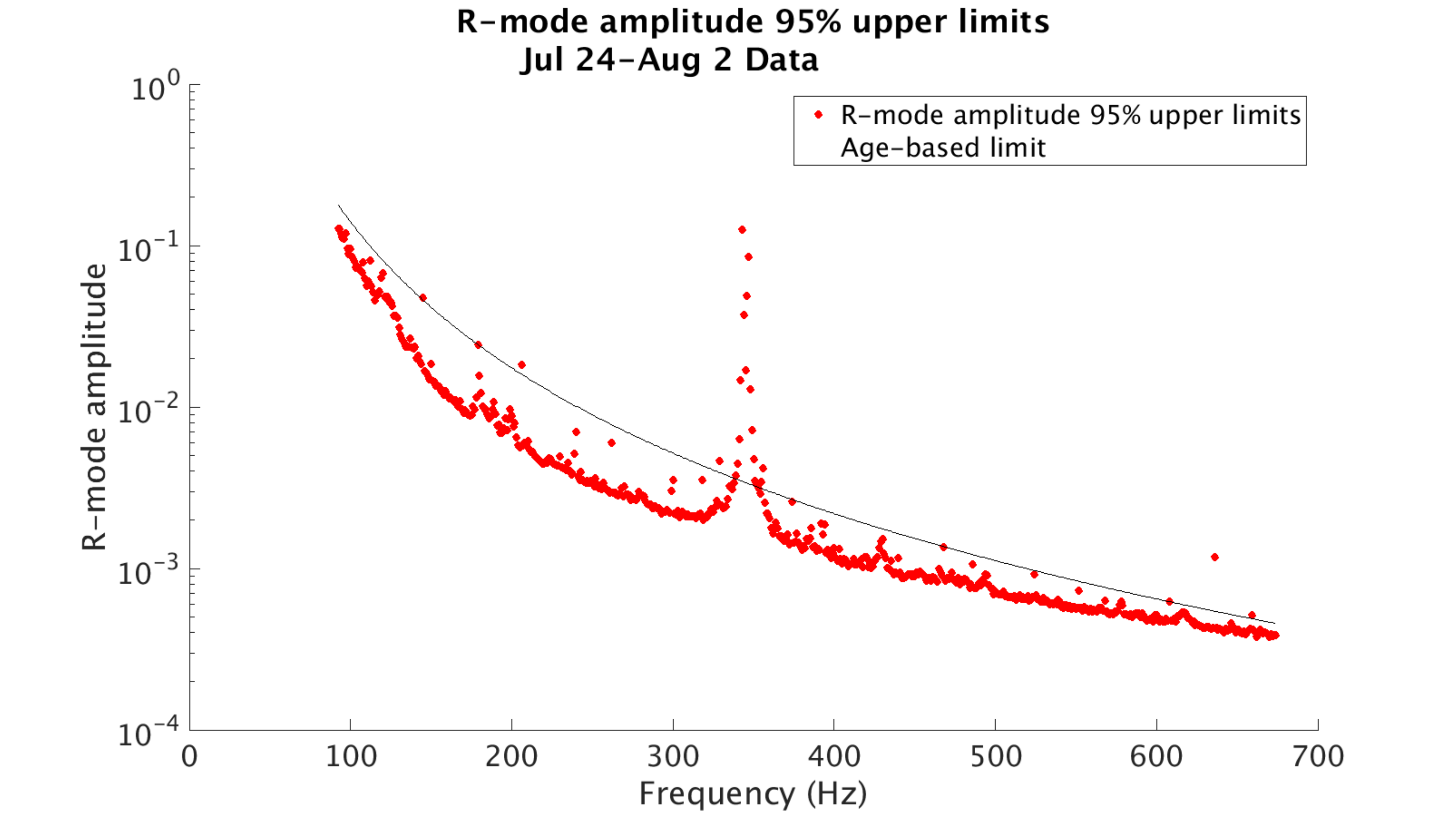}
  \caption{Upper limits at 95\% confidence (red circles) compared to the initial age-based limit on $r$-mode amplitude $\alpha $ (black line).}
  \label{RMode}
\end{figure}

Figure ~\ref{ULE} shows the 95\% confidence upper limits (ULs) over the full band for the July-August data set, which was the more sensitive of the two because of its much greater livetime (642 SFTs vs. 374 SFTs for the October data set).  The blue curve represents the expected sensitivity of the search for this data set, computed from the power spectral density at each frequency; there is good agreement with the ULs. The black line represents the age-based limit derived when first considering the parameter space. Its intersection with the ULs at either end of the plot is a confirmation that we correctly estimated the frequency band to search over.

Figure ~\ref{Ellip} is a similar plot converting the upper limits on $h_{0}$ to upper limits on fiducial ellipticity $\epsilon $, using the formula \cite{Saulson1994}

\begin{equation}
\label{EllipForm}
\epsilon = \frac{c^{4}}{4\pi ^{2}G}\frac{d}{I_{zz}f^{2}}h_{0}.
\end{equation}

\noindent The black curve represents the age-based limit on fiducial ellipticity, using the same assumptions (braking index $n = 5$, age $\tau = 300$ years) used in the parameter space calculations.

The amplitude $\alpha $ of $r$-mode oscillations in a neutron star is related to the gravitational wave strain amplitude $h_{0}$ by \cite{OwenRModes}

\begin{equation}
\label{RModeForm}
\alpha = 0.028\left ( \frac{h_{0}}{10^{-24}} \right )\left ( \frac{r}{1 ~\textup{kpc}} \right )\left ( \frac{100 ~\textup{Hz}}{f} \right )^{3}.
\end{equation}

Figure ~\ref{RMode} uses this formula to convert the upper limits on $h_{0}$ to upper limits on the $r$-mode amplitude $\alpha $. The black curve represents the age-based limit on $\alpha $, under the same age assumptions used for $h_{0}$ and ellipticity, but with $n = 7$, which characterizes $r$-mode emission. In all three plots, the age-based limit is beaten everywhere the upper limits lie below the black curve. 

\section{Discussion}
\label{s:disc}
This search has placed the first explicit upper limits on continuous gravitational wave strength from the nearby globular cluster NGC~6544 for spin-down ages as young as 300 years, and is the first directed CW search for any globular cluster.  The most stringent upper limits on strain ($h_{0}^{\textup{UL}}$) obtained were $h_{0}^{\textup{UL}} = 6.7 \times 10^{-25}$ for the 173-174 Hz band in the October data set, and $h_{0}^{\textup{UL}} = 6.0 \times 10^{-25}$ for the 170-171 Hz band in the July-August data. 

The best upper limit is comparable to the best upper limit of $7 \times 10^{-25}$ at $150$ Hz obtained by the Cas~A search \cite{S5CasA}; the recent search over nine supernova remnants, done without resampling, \cite{S6CasAFriends} set upper limits as low as $3.7 \times 10^{-25}$ for the supernova remnant G93.3+6.9, but used a coherence time of over 23 days (and a frequency band of only 264 Hz). The same analysis set a comparable 95\% upper limit of $6.4 \times 10^{-25}$ for the supernova remnant G1.9+0.3, as expected given its similar declination and the search's similar coherence time (9.2 days for NGC~6544 vs. 9.1 days for G1.9+0.3). Note, however, that the G1.9+0.3 search was limited to a 146 Hz search band, compared to 583 Hz for NGC~6544. The search reported here was carried out at substantially less computational cost because of barycentric resampling, and could thus search over a much larger parameter space.

The best upper limit on fiducial ellipticity, established using the July-August data set, was $\epsilon = 8.5 \times 10^{-6}$, for the $1$-Hz band starting at $670$ Hz. This is comparable to the best upper limit ($4 \times 10^{-5}$) obtained by the Cas~A search \cite{Wette2009}; the supernova remnant search \cite{S6CasAFriends} set a comparable upper limit on fiducial ellipticity at $7.6 \times 10^{-5}$ for the supernova remnant G1.9+0.3.

These ellipticities are within the range of maximum theoretical ellipticities predicted for stars with some exotic phases in the core \cite{OwenStrangeStars,OwenJ-McD2013}, and the lowest of them is achievable for purely nucleonic stars with a sufficiently stiff equation of state and low mass \cite{OwenJ-McD2013}. Hence the search could have detected some exotic stars if they were supporting close to their maximum possible ellipticity; however, the lack of a detection cannot be used to infer constraints on the composition of any star, since the deformation could be much less than its maximum supportable value.

The first observing run of the Advanced LIGO detectors began in September 2015 \cite{Prospects} and the sensitivity of the detectors is already three times or more better than that used in this search, with an order of magnitude improvement over \ac{S6} expected eventually \cite{AdvLIGO}. The barycentric resampling algorithm first implemented in this search is being used extensively in CW searches in the advanced detector era; it has been integrated into the search codes for both coherent searches (like the supernova remnant search) \cite{IdrisyThesis} and semi-coherent searches (Einstein@Home).

\section{Acknowledgments}
The authors gratefully acknowledge the support of the United States
National Science Foundation (NSF) for the construction and operation of the
LIGO Laboratory and Advanced LIGO as well as the Science and Technology Facilities Council (STFC) of the
United Kingdom, the Max-Planck-Society (MPS), and the State of
Niedersachsen/Germany for support of the construction of Advanced LIGO 
and construction and operation of the GEO600 detector. 
Additional support for Advanced LIGO was provided by the Australian Research Council.
The authors gratefully acknowledge the Italian Istituto Nazionale di Fisica Nucleare (INFN),  
the French Centre National de la Recherche Scientifique (CNRS) and
the Foundation for Fundamental Research on Matter supported by the Netherlands Organisation for Scientific Research, 
for the construction and operation of the Virgo detector
and the creation and support  of the EGO consortium. 
The authors also gratefully acknowledge research support from these agencies as well as by 
the Council of Scientific and Industrial Research of India, 
Department of Science and Technology, India,
Science \& Engineering Research Board (SERB), India,
Ministry of Human Resource Development, India,
the Spanish Ministerio de Econom\'ia y Competitividad,
the Conselleria d'Economia i Competitivitat and Conselleria d'Educaci\'o, Cultura i Universitats of the Govern de les Illes Balears,
the National Science Centre of Poland,
the European Commission,
the Royal Society, 
the Scottish Funding Council, 
the Scottish Universities Physics Alliance, 
the Hungarian Scientific Research Fund (OTKA),
the Lyon Institute of Origins (LIO),
the National Research Foundation of Korea,
Industry Canada and the Province of Ontario through the Ministry of Economic Development and Innovation, 
the Natural Science and Engineering Research Council Canada,
Canadian Institute for Advanced Research,
the Brazilian Ministry of Science, Technology, and Innovation,
Funda\c{c}\~ao de Amparo \`a Pesquisa do Estado de S\~ao Paulo (FAPESP),
Russian Foundation for Basic Research,
the Leverhulme Trust, 
the Research Corporation, 
Ministry of Science and Technology (MOST), Taiwan
and
the Kavli Foundation.
The authors gratefully acknowledge the support of the NSF, STFC, MPS, INFN, CNRS and the
State of Niedersachsen/Germany for provision of computational resources.

This document has been assigned LIGO Laboratory document number \texttt{LIGO-P1500225}.

\bibliography{S6GlobClust}

\end{document}